\documentclass[a4paper,fleqn,usenatbib]{mnras}
\usepackage[normalem]{ulem}

\usepackage{newtxtext,newtxmath}

\usepackage[T1]{fontenc}
\usepackage{ae,aecompl}

\usepackage{graphicx}	
\usepackage{amsmath}	
\usepackage{amssymb}	
\usepackage{pdflscape}
\usepackage{footmisc}

\defcitealias{2015AJ....149...91P}{Paper\,I}
\defcitealias{2015MNRAS.447..927M}{Paper\,II}
\defcitealias{2015MNRAS.451..312N}{Paper\,IV}
\defcitealias{2017MNRAS.464.3636M}{Paper\,IX}
\defcitealias{2016MNRAS.463..449S}{Paper\,X}
\defcitealias{2015MNRAS.454.4197R}{Paper\,V}
\defcitealias{2017AJ....153...19S}{Paper\,VIII}
\defcitealias{2018MNRAS.tmp..176S}{Paper\,XIII}

\title[Public Data Release]{The {\it Hubble Space Telescope} UV Legacy
  Survey of Galactic Globular Clusters - XVII.  Public Catalogue
  Release.\thanks{Based on observations with the NASA/ESA {\it Hubble
      Space Telescope}, obtained at the Space Telescope Science
    Institute, which is operated by AURA, Inc., under NASA contract
    NAS 5-26555.}\thanks{All of the data products are available at
    MAST as a High Level Science Product under the project HUGS:
    \url{https://archive.stsci.edu/prepds/hugs/}}}
\author[Nardiello et al.]{D.\ Nardiello$^{1,2}$\thanks{E-mail: domenico.nardiello@unipd.it}, 
M.\ Libralato$^{3}$, 
G.\ Piotto$^{1,2}$,
J.\ Anderson$^{3}$,
A.\ Bellini$^{3}$,
\newauthor
A.\ Aparicio$^{4,5}$,
L.\ R.\ Bedin$^{2}$,
S.\ Cassisi$^{6}$,
V.\ Granata$^{1,2}$,
I.\ R.\ King$^{7}$,
F.\ Lucertini$^{1}$,
\newauthor
A.\ F.\ Marino$^{8}$,
A.\ P.\ Milone$^{1}$,
S.\ Ortolani$^{1}$,
I.\ Platais$^{9}$,
R.\ P.\ van der Marel$^{3,9}$\\
$^{1}$Dipartimento di Fisica e Astronomia ``Galileo Galilei'', Universit\`a di Padova, Vicolo dell'Osservatorio 3, Padova IT-35122 \\
$^{2}$Istituto Nazionale di Astrofisica - Osservatorio Astronomico di Padova, Vicolo dell'Osservatorio 5, Padova, IT-35122 \\
$^{3}$Space Telescope Science Institute, 3800 San Martin Drive, Baltimore, MD 21218, USA \\
$^{4}$Instituto de Astrofisica de Canarias, E-38200 La Laguna, Tenerife, Canary Islands, Spain\\
$^{5}$Department of Astrophysics, University of La Laguna, E-38200 La Laguna, Tenerife, Canary Islands, Spain\\
$^{6}$Osservatorio Astronomico d'Abruzzo, Via M. Maggini sn., I-64100 Teramo, Italy \\
$^{7}$Department of Astronomy, University of Washington, Box 351580, Seattle, WA 98195, USA \\
$^{8}$Research School of Astronomy and Astrophysics, The Australian National University, Cotter Road, Weston, ACT, 2611, Australia \\
$^{9}$Center for Astrophysical Sciences, Department of Physics \& Astronomy, Johns Hopkins University, Baltimore, MD 21218, USA\\
}

\date{Accepted 2018 September 11. Received 2018 September 10; in original form 2018 June 27}

\pubyear{2018}

\begin{document}
\label{firstpage}
\pagerange{\pageref{firstpage}--\pageref{lastpage}}
\maketitle

\begin{abstract}
In this paper we present the astro-photometric catalogues of 56
globular clusters and one open cluster. Astrometry and photometry are
mainly based on images collected within the ``HST Legacy Survey of
Galactic Globular Clusters: Shedding UV Light on Their Populations and
Formation'' (GO-13297, PI:~Piotto), and the ``ACS Survey of Galactic
Globular Clusters'' (GO-10775, PI:~Sarajedini). For each source in the
catalogues for which we have reliable proper motion we also publish a
membership probability for separation of field and cluster
stars. These new catalogues, which we make public in Mikulski Archive
for Space Telescopes, replace previous catalogues by Paper\,VIII of this series.
\end{abstract}




\section{Introduction}
\label{sec:intro}
Nowadays the presence of multiple stellar populations (MPs) in
globular clusters (GCs) is a commonly accepted observational fact,
even though our understanding of their origin is still far from satisfying
(\citealt{2015MNRAS.454.4197R,2015arXiv151001330B}, \citealt{2017arXiv171201286B}).
The ``HST Legacy Survey of Galactic Globular Clusters: Shedding UV
Light on Their Populations and Formation'' (GO-13297, PI:~Piotto)
observations, combined with the optical data from the ``ACS Survey of
Galactic Globular Clusters'' (ACS\,GCS; GO-10775, PI:~Sarajedini) have
provided key building blocks for the observational edifice of MPs.
These datasets have allowed us to demonstrate their ubiquitous
presence in all Galactic GCs studied in enough details,
convincingly showing the existence of discrete populations,
establishing a tight connection between photometric and spectroscopic
data, and spurring further studies by discovering populations with
particularly complex chemical patterns (\citealt[hereafter
  Paper\,I]{2015AJ....149...91P};
\citealt{2017MNRAS.464.3636M,2018ApJ...859...81M} and references
therein).

In this paper, we present and publish the final catalogues.  These
catalogues contain astrometric positions, F275W, F336W, F438W, F606W,
and F814W photometry and cluster membership from proper motions (PMs)
of stars in the central regions of 56 GCs and the old super metal-rich
open cluster (OC) NGC\,6791, presented in
\citetalias{2015AJ....149...91P}.  The GO-13297 data are complemented
here by the Wide Field Camera 3 (WFC3) images collected within the
GO-12311 (PI:~Piotto) and GO-12605 (PI:~Piotto) programs, used as
pilot projects for the more extended UV Legacy survey.  As discussed
in Section~\ref{sec:newdr}, the catalogues presented in this paper
replace our preliminary catalogues published by
\citet[Paper\,VIII]{2017AJ....153...19S}.  The complete GO-13297
dataset also includes the astrometry and photometry catalogues of the
external fields taken with the Advanced Camera for Surveys (ACS), in
parallel with the GO-13297 WFC3/UVIS central fields and published by
\citet[Paper\,XIII]{2018MNRAS.tmp..176S}.

The paper is organised as follows. Section~\ref{sec:obs} is dedicated
to the observations and data reduction; section~\ref{sec:cmds} briefly
presents the colour-magnitude diagrams; the proper motion measurements and
the methodology to estimate membership probability are described in
section~\ref{sec:rpm}. 
Section~\ref{sec:newdr} discusses the improvements of the new data
reduction with respect to the preliminary one of
\citetalias{2017AJ....153...19S}.
In section~\ref{sec:dr} we describe the content 
of the data release tables.

\section{Observations and data reduction}
\label{sec:obs}

In this paper, we present high-precision stellar astrometry and
photometry from WFC3/UVIS and ACS/WFC observations of 56 GCs and 
the old open cluster NGC~6791.  The GCs were all observed with ACS/WFC 
in F606W and F814W bands within GO-10775 (PI:\, A.~Sarajedini). For the 
open cluster NGC\,6791 we used the ACS/WFC data in the same filters
collected within GO-10265 (PI:\,T.~Brown). Observations in the UV/blue
{\it HST} bands (F275W, F336W, and F438W) of 55 clusters were
collected with the WFC3/UVIS camera within GO-12311 (PI:\,G.~Piotto),
GO-12605 (PI:\,G.~Piotto) , and GO-13297 (PI:\,G.~Piotto) programs. A
complete log of these observations is presented in
\citetalias{2015AJ....149...91P}.  In addition to the data used in
\citetalias{2015AJ....149...91P}, for NGC\,0104 we also incorporated 
F336W observations from GO-11729 (PI:\,J.~Holtzman) and GO-12971
(PI:\,H.~Richer), and F435W images collected with ACS/WFC within
GO-9443 (PI:\,I.~King) and GO-9281 (PI:\,J.~Grindlay). For NGC\,6752
we used F275W data from GO-12311, F336W images from GO-11729, and
F435W ACS/WFC observations obtained by GO-12254 (PI:\,A.~Cool).

\subsection{First-pass photometry}
\label{ssec:1stp}
We worked on \texttt{\_flc} images, which are \texttt{\_flt} exposures
corrected for charge-transfer efficiency (CTE) defects
(\citealt{2010PASP..122.1035A}). For the data reduction we used an
evolution of the software described in \citet{2008AJ....135.2055A}. A
detailed description of the adopted tools is given by
\citet{2017ApJ...842....6B}, \citet{2018MNRAS.477.2004N}, and
\citet{2018arXiv180505332L}.

\begin{figure}
\includegraphics[width=0.5\textwidth]{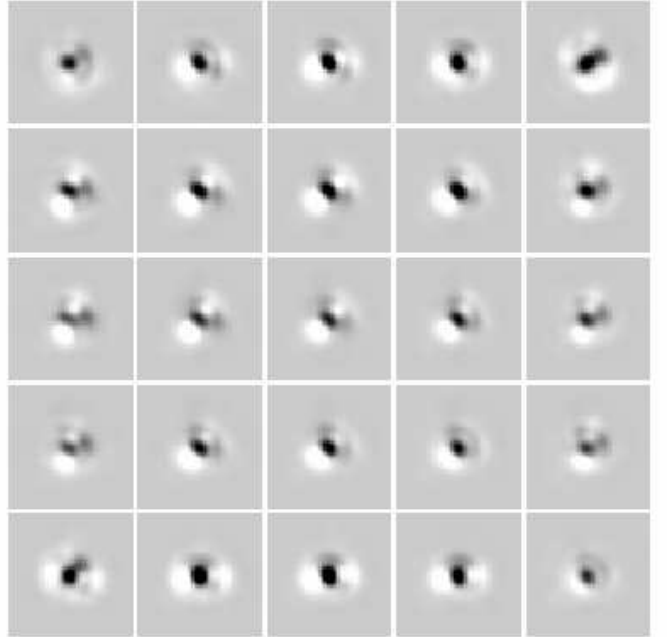}
\caption{ A $5 \times 5$ grid of perturbation PSFs that modify the
  library PSF array in the case of the image \texttt{j9l959f6q}
  (NGC\,6352). The total variation across the the grid goes from $\sim
  -2$\% to $\sim 1$\% of the star's total flux.
\label{fig:1}}
\end{figure}

Briefly, for each image, we accounted on the spatial and time
dependence of the Point Spread Function (PSF) by constructing an
optimal PSF for each exposure by perturbing the "library"
PSF\footnote{http://www.stsci.edu/$\sim$jayander/STDPSFs/} appropriate
for each filter.  In order to obtain the perturbed PSFs we used the
\texttt{FORTRAN} program \texttt{hst1pass} (see also
\citealt{2018ApJ...853...86B}); we selected bright, unsaturated,
isolated stars, we measured the flux and the positions using the
library PSFs, and finally we subtracted a model of each star to the
real star. The residuals of the subtraction are averaged to form a
grid of residuals used to perturb the library PSFs. This grid has
dimensions that go from $1 \times 1$ to $5 \times 5$, depending on the
total number of stars available in the field. Each element of the grid
corresponds to a different fiducial location on the detector, and we
used linear interpolation to evaluate the PSF between these locations.
Nine rounds of iterations allowed us to arrive at an evenly spaced set
of perturbation PSFs from the random distribution of stars in each
image.  An example of the grid of residual PSFs is shown in
Fig.~\ref{fig:1}.

With these arrays of PSFs, we extracted the preliminary catalogues
using the program \texttt{hst1pass}. This program measures positions
and fluxes of the stars on the single HST exposures, without performing
any neighbour subtraction.  It is even able to make measurements 
of saturated stars, using the technique described in
\citet{2004acs..rept...17G} and \citet{2010wfc..rept...10G}.
We corrected the positions of the stars for geometric
distortion using the routines described in
\citet[ACS/WFC]{2006acs..rept....1A}, and \citet{2009PASP..121.1419B},
\citet[WFC3/UVIS]{2011PASP..123..622B}.

We transformed all the catalogues into a common reference system. We
adopted the {\it Gaia} Data Release 1 catalogue
(\citealt{2016A&A...595A...2G}) as the reference system for
positions. In this way, the {\texttt X}- and {\texttt Y}-axes are
aligned with West and North, respectively.  We de-projected the {\it
  Gaia} ($\alpha$, $\delta$)-coordinates onto a tangent plane with the
cluster centre from \citet{2010AJ....140.1830G} as the tangent point
($\alpha_0$, $\delta_0$).  We transformed these coordinates into
WFC3/UVIS pixels (pixel scale 0.0395\,arcsec\,pixel$^{-1}$;
\citealt{2018wfci.book.....D}), positioning the centre of the cluster
in the pixel (5000,5000).  In the first iteration, we found the
six-parameter linear transformations between this master catalogue and
each of the single-exposure catalogues by using unsaturated, bright,
and isolated stars, and used this transformation to transform all
stars measured in each exposure into this reference frame.  We
collated these lists and extracted a new master catalogue with the
3$\sigma$-clipped average stellar positions.  We then used this new
catalogue to improve the transformations, iterating until the
precision on the transformed positions did not improve.  For each
filter, the photometric zero-point of each individual catalogue is
tied to those of the deepest exposure. For each filter, we
obtained a final catalogue containing the 3$\sigma$-clipped average
stellar positions and magnitudes in that filter (``first-pass''
photometry, similar to that used in \citetalias{2015AJ....149...91P}).

\subsection{Multiple-pass photometry}
In the ``multiple-pass'' photometry stage, we used the images, 
the PSF arrays and the transformations obtained during the ``first-pass'' 
to simultaneously find and measure stars in all the individual
exposures. The tool we used is the \texttt{FORTRAN} software
\texttt{kitchen\_sync2} (KS2;
\citealt{2017ApJ...842....6B,2018MNRAS.477.2004N,2018arXiv180505332L}).
The software analyses all the images simultaneously to find and
measure the sources; in this way it is able to also measure the stars
that cannot be detected in individual exposures. To avoid spurious
detection caused by artifacts of the PSF, and in order to detect
faint stars close to bright stars, KS2 creates an ad-hoc mask for
bright and saturated stars, which were included in the one-pass 
catalogues.

The finding procedure is accomplished through different iterations. During the
first iteration, the software identifies the bright stars and
subtracts them. In the following step, the routine searches for stars
that are fainter than the stars from previous iteration and then measures and
subtracts them. In each iteration we defined different criteria (which
are increasingly more relaxed from the first to the last iteration) to qualify
a source as a star.  We chose to iterate 8 times: in the first five
iterations we required that a stars be present both in the F606W and
F814W exposures; in iteration six, seven, and eight we performed the
finding on the F275W, F336W, and F438W exposures to detect the stars
that are brighter in the UV/blue filters than optical filters (i.e., white
dwarfs) and not detected in optical filters.

The KS2 software generated astrometric and photometric catalogues of
stars using three different methods. A detailed description of the
three methods is given in \citet{2017ApJ...842....6B}.

Method-1 gives the best results for stars that are bright enough to 
generate star-like profiles in individual exposures.  During the
finding stage, the routine searches for a distinct peak in a $5 \times
5$ pixel$^2$ raster and measures, in each image, the flux and the position
of the source using an appropriate local PSF, after subtracting the
neighbour stars.  The local sky value is computed inside an annulus
with an inner radius $r_{\rm in}=4$\,pixels and outer radius $r_{\rm
  out}=8$\,pixels. The final flux and position of a star in a filter
is given by a robust average of the fluxes and positions measured in
the single exposures.

Method-2 works well for faint stars and crowded
environments. Starting from the position obtained during the finding
stage, KS2 performs weighted aperture photometry of the star in a
raster of $5 \times 5$ pixels, after neighbour-subtraction; each pixel
is weighted in such a way that pixel containing neighbour stars are
down-weighted.  The local sky is computed as above for method-1. The
final flux is a robust average of the fluxes obtained in the single
exposures.

Method-3 works well in very crowded environments. It is similar to
method-2, but uses only the pixels inside a radius $r=0.75$\,pixel
from the centre of the star and the sky is calculated in a tight
annulus with $r_{\rm in}=2$\,pixels and $r_{\rm out}=4$\,pixels.

Saturated stars are not measured by KS2.  They were, however, included
in the one-pass based catalogues using techniques described above.

In addition to the astro-photometric catalogue, KS2 also outputs 
stacked versions of the fields obtained from the \texttt{\_flc} 
images.  Excluding NGC\,0104, NGC\,6752, and NGC\,5897, for all the clusters we
generated 7 different stacked images: one for the filters F275W,
F336W, and F438W, and two for the filters F606W and F814W,
separating short- and long-exposure images. For
NGC\,5897 the F814W short-exposure  image is unusable. For
NGC\,0104 and NGC\,6752 we generated 8 stacked images: one for the
filters F275W and F336W, and two for the filters F435W, F606W and
F814W, one for short-exposure observations and one for the long
ones.

\begin{table*}
  \caption{Difference between $m_{\rm CAL,{\rm X}}$ and
    $m^{flc}_{\rm PSF, X}$ for each filter X }
    \label{tab1}
    \begin{tabular}{ l c c c c c}
\hline
\multicolumn{1}{c}{Cluster} &
\multicolumn{1}{c}{F275W} &
\multicolumn{1}{c}{F336W} &
\multicolumn{1}{c}{F438W$^1$} &
\multicolumn{1}{c}{F606W} &
\multicolumn{1}{c}{F814W} \\ 
\hline
NGC\,0104 &  28.75$\pm$0.04 &  30.42$\pm$0.02 &  30.71$\pm$0.01 &  30.60$\pm$0.03 &  29.70$\pm$0.03  \\
NGC\,0288 &  28.94$\pm$0.01 &  29.66$\pm$0.01 &  28.85$\pm$0.01 &  31.62$\pm$0.01 &  30.88$\pm$0.01  \\
NGC\,0362 &  29.19$\pm$0.04 &  29.66$\pm$0.02 &  29.14$\pm$0.01 &  31.80$\pm$0.02 &  31.04$\pm$0.02  \\
NGC\,1261 &  29.75$\pm$0.02 &  29.84$\pm$0.01 &  30.38$\pm$0.01 &  32.71$\pm$0.02 &  31.85$\pm$0.02  \\
NGC\,1851 &  30.21$\pm$0.02 &  29.94$\pm$0.01 &  30.18$\pm$0.01 &  32.71$\pm$0.02 &  31.82$\pm$0.02  \\
NGC\,2298 &  30.34$\pm$0.03 &  29.66$\pm$0.01 &  30.14$\pm$0.01 &  32.71$\pm$0.02 &  31.82$\pm$0.02  \\
NGC\,2808 &  29.89$\pm$0.02 &  30.33$\pm$0.02 &  29.77$\pm$0.02 &  32.74$\pm$0.03 &  31.87$\pm$0.05  \\
NGC\,3201 &  29.62$\pm$0.01 &  29.53$\pm$0.01 &  29.39$\pm$0.01 &  31.35$\pm$0.01 &  30.46$\pm$0.02  \\
NGC\,4590 &  29.53$\pm$0.01 &  29.51$\pm$0.01 &  29.26$\pm$0.01 &  31.64$\pm$0.02 &  30.89$\pm$0.02  \\
NGC\,4833 &  29.85$\pm$0.01 &  29.66$\pm$0.01 &  30.15$\pm$0.01 &  31.78$\pm$0.01 &  31.03$\pm$0.01  \\
NGC\,5024 &  30.58$\pm$0.01 &  29.96$\pm$0.01 &  30.46$\pm$0.01 &  32.68$\pm$0.03 &  31.78$\pm$0.03  \\
NGC\,5053 &  29.74$\pm$0.03 &  29.84$\pm$0.01 &  30.34$\pm$0.01 &  32.68$\pm$0.01 &  31.82$\pm$0.02  \\
NGC\,5272 &  28.96$\pm$0.02 &  29.66$\pm$0.02 &  28.87$\pm$0.01 &  31.64$\pm$0.02 &  30.90$\pm$0.03  \\
NGC\,5286 &  29.66$\pm$0.02 &  29.56$\pm$0.02 &  29.33$\pm$0.01 &  32.71$\pm$0.03 &  31.85$\pm$0.03  \\
NGC\,5466 &  30.14$\pm$0.03 &  29.65$\pm$0.01 &  30.13$\pm$0.01 &  32.69$\pm$0.02 &  31.82$\pm$0.02  \\
NGC\,5897 &  29.84$\pm$0.01 &  29.65$\pm$0.01 &  30.18$\pm$0.01 &  32.68$\pm$0.02 &  31.82$\pm$0.02  \\
NGC\,5904 &  29.51$\pm$0.01 &  29.51$\pm$0.01 &  29.22$\pm$0.01 &  31.71$\pm$0.02 &  30.82$\pm$0.02  \\
NGC\,5927 &  29.58$\pm$0.04 &  29.51$\pm$0.01 &  29.80$\pm$0.02 &  32.71$\pm$0.02 &  31.85$\pm$0.02  \\
NGC\,5986 &  29.61$\pm$0.01 &  29.49$\pm$0.01 &  29.33$\pm$0.01 &  32.72$\pm$0.02 &  31.82$\pm$0.03  \\
NGC\,6093 &  29.78$\pm$0.02 &  30.31$\pm$0.02 &  29.60$\pm$0.01 &  32.69$\pm$0.03 &  31.79$\pm$0.03  \\
NGC\,6101 &  29.86$\pm$0.01 &  29.84$\pm$0.01 &  30.35$\pm$0.01 &  32.77$\pm$0.02 &  31.91$\pm$0.02  \\
NGC\,6121 &  29.70$\pm$0.01 &  29.49$\pm$0.01 &  29.36$\pm$0.01 &  29.84$\pm$0.02 &  29.14$\pm$0.02  \\
NGC\,6144 &  29.63$\pm$0.02 &  29.51$\pm$0.01 &  29.28$\pm$0.01 &  32.68$\pm$0.01 &  31.82$\pm$0.02  \\
NGC\,6171 &  30.13$\pm$0.04 &  29.65$\pm$0.01 &  30.17$\pm$0.01 &  31.64$\pm$0.02 &  30.90$\pm$0.02  \\
NGC\,6205 &  29.00$\pm$0.02 &  29.66$\pm$0.01 &  28.97$\pm$0.01 &  31.72$\pm$0.03 &  30.83$\pm$0.03  \\
NGC\,6218 &  29.67$\pm$0.01 &  29.51$\pm$0.01 &  29.36$\pm$0.01 &  31.24$\pm$0.02 &  30.34$\pm$0.02  \\
NGC\,6254 &  29.68$\pm$0.01 &  29.51$\pm$0.01 &  29.37$\pm$0.01 &  31.23$\pm$0.02 &  30.34$\pm$0.02  \\
NGC\,6304 &  29.55$\pm$0.03 &  29.51$\pm$0.02 &  29.28$\pm$0.01 &  32.68$\pm$0.02 &  31.82$\pm$0.02  \\
NGC\,6341 &  29.55$\pm$0.02 &  29.51$\pm$0.02 &  29.24$\pm$0.02 &  31.72$\pm$0.02 &  30.90$\pm$0.03  \\
NGC\,6352 &  29.58$\pm$0.02 &  29.53$\pm$0.01 &  29.22$\pm$0.01 &  31.72$\pm$0.02 &  30.90$\pm$0.02  \\
NGC\,6362 &  29.58$\pm$0.02 &  29.57$\pm$0.01 &  29.40$\pm$0.01 &  31.64$\pm$0.01 &  30.90$\pm$0.02  \\
NGC\,6366 &  29.73$\pm$0.05 &  29.51$\pm$0.02 &  29.29$\pm$0.01 &  31.72$\pm$0.02 &  30.83$\pm$0.02  \\
NGC\,6388 &  30.16$\pm$0.03 &  29.65$\pm$0.05 &  30.11$\pm$0.02 &  32.68$\pm$0.05 &  31.81$\pm$0.06  \\
NGC\,6397 &  29.63$\pm$0.01 &  29.53$\pm$0.01 &  29.36$\pm$0.01 &  29.29$\pm$0.02 &  28.40$\pm$0.02  \\
NGC\,6441 &  30.29$\pm$0.07 &  29.64$\pm$0.04 &  30.06$\pm$0.02 &  32.68$\pm$0.05 &  31.81$\pm$0.06  \\
NGC\,6496 &  29.67$\pm$0.07 &  29.50$\pm$0.01 &  29.47$\pm$0.01 &  32.68$\pm$0.02 &  31.82$\pm$0.02  \\
NGC\,6535 &  29.68$\pm$0.03 &  29.51$\pm$0.01 &  29.30$\pm$0.01 &  31.64$\pm$0.01 &  30.90$\pm$0.02  \\
NGC\,6541 &  29.63$\pm$0.02 &  29.50$\pm$0.02 &  29.36$\pm$0.02 &  31.72$\pm$0.02 &  30.90$\pm$0.03  \\
NGC\,6584 &  29.55$\pm$0.03 &  29.53$\pm$0.01 &  29.29$\pm$0.01 &  32.72$\pm$0.02 &  31.85$\pm$0.02  \\
NGC\,6624 &  29.67$\pm$0.03 &  29.48$\pm$0.02 &  29.58$\pm$0.01 &  32.71$\pm$0.02 &  31.81$\pm$0.02  \\
NGC\,6637 &  29.86$\pm$0.02 &  29.66$\pm$0.01 &  30.10$\pm$0.01 &  32.68$\pm$0.02 &  31.78$\pm$0.02  \\
NGC\,6652 &  29.67$\pm$0.05 &  29.53$\pm$0.01 &  29.65$\pm$0.01 &  32.68$\pm$0.02 &  31.79$\pm$0.02  \\
NGC\,6656 &  29.69$\pm$0.02 &  29.99$\pm$0.01 &  30.19$\pm$0.01 &  30.70$\pm$0.02 &  29.99$\pm$0.02  \\
NGC\,6681 &  29.69$\pm$0.01 &  29.48$\pm$0.01 &  29.62$\pm$0.01 &  31.72$\pm$0.02 &  30.90$\pm$0.02  \\
NGC\,6715 &  30.51$\pm$0.03 &  29.97$\pm$0.04 &  30.51$\pm$0.02 &  32.69$\pm$0.04 &  31.82$\pm$0.04  \\
NGC\,6717 &  29.51$\pm$0.01 &  29.54$\pm$0.01 &  29.26$\pm$0.01 &  31.63$\pm$0.02 &  30.90$\pm$0.02  \\
NGC\,6723 &  29.59$\pm$0.02 &  29.54$\pm$0.01 &  29.27$\pm$0.01 &  31.73$\pm$0.02 &  30.91$\pm$0.02  \\
NGC\,6752 &  28.84$\pm$0.02 &  30.05$\pm$0.01 &  32.14$\pm$0.01 &  30.21$\pm$0.02 &  29.46$\pm$0.02  \\
NGC\,6779 &  29.69$\pm$0.01 &  29.47$\pm$0.01 &  29.58$\pm$0.01 &  32.68$\pm$0.02 &  31.82$\pm$0.02  \\
NGC\,6791 &  29.49$\pm$0.05 &  29.48$\pm$0.01 &  29.45$\pm$0.01 &  30.60$\pm$0.01 &  29.72$\pm$0.01  \\
NGC\,6809 &  29.71$\pm$0.01 &  29.47$\pm$0.01 &  29.36$\pm$0.01 &  30.97$\pm$0.02 &  30.22$\pm$0.02  \\
NGC\,6838 &  29.68$\pm$0.02 &  29.50$\pm$0.01 &  29.35$\pm$0.01 &  31.04$\pm$0.02 &  30.21$\pm$0.02  \\
NGC\,6934 &  29.67$\pm$0.02 &  29.50$\pm$0.01 &  29.42$\pm$0.01 &  32.69$\pm$0.02 &  31.79$\pm$0.02  \\
NGC\,6981 &  29.55$\pm$0.02 &  29.51$\pm$0.01 &  29.43$\pm$0.01 &  31.64$\pm$0.02 &  30.91$\pm$0.02  \\
NGC\,7078 &  29.53$\pm$0.02 &  29.65$\pm$0.02 &  29.34$\pm$0.01 &  31.64$\pm$0.03 &  30.90$\pm$0.04  \\
NGC\,7089 &  29.68$\pm$0.02 &  29.54$\pm$0.02 &  29.29$\pm$0.02 &  32.69$\pm$0.03 &  31.79$\pm$0.04  \\
NGC\,7099 &  29.58$\pm$0.02 &  29.50$\pm$0.01 &  29.35$\pm$0.01 &  31.72$\pm$0.02 &  30.82$\pm$0.02  \\
\hline
\end{tabular}

$^1$ For NGC\,0104 and NGC\,6752, the value is referred to ACS/WFC F435W filter.

\end{table*}

\subsection{Photometric Calibration}

We calibrated the output photometry from KS2 into the Vega-mag system
by comparing aperture photometry on \texttt{\_drc} images (which are
normalised to the exposure time of 1\,s) with our PSF-based photometry.

For aperture photometry on \texttt{\_drc} images we used an aperture
radius $r_{\rm AP}=0.2$\,arcsec.  We adopted the $r_{\rm AP}$ after
testing different apertures (from 0.04\,arcsec to 0.4\,arcsec) on a
sample of 10 GCs images with different crowding level and total number
of stars. We found that $r_{\rm AP}=0.2$\,arcsec gives, on average,
the lowest error on the determination of the zeropoint that converts 
the instrumental magnitudes to calibrated magnitudes. It represents a 
fair compromise between measuring as much flux as possible for the stars 
($\gtrsim 80$\,\%) on the \texttt{\_drc} images and avoiding the 
contamination from neighbour stars.

We corrected the aperture photometry magnitudes $m^{\rm drc}_{\rm AP}$
with the appropriate aperture correction\footnote{For ACS we used the aperture
  correction tabulated by \citet{2016AJ....152...60B}, while for WFC3
  we used the corrections in \citet{2017wfc..rept...14D}}, obtaining
the magnitudes for an infinite-aperture radius $m^{\rm drc}_{\rm AP, \infty}$.

For each filter we cross-identified the stars in common between the
catalogue obtained using KS2 and that obtained from aperture photometry
on \texttt{\_drc} images, and computed the 3$\sigma$-clipped
average $<\delta m>$, and the associated error, of the difference
$\delta m = m^{\rm drc}_{\rm AP, \infty}-m^{flc}_{\rm PSF}$, where
$m^{flc}_{\rm PSF}$ is the magnitude associated with the PSF photometry
on \texttt{\_flc} images output of KS2.

The calibrated magnitude $m_{\rm CAL,{\rm X}}$ of a star in the filter X
is given by:
\begin{equation}
m_{\rm CAL,{\rm X}}=m^{flc}_{\rm PSF,{\rm X}}+<\delta m> + {\rm ZP}_{\rm X}
\end{equation}
 where ZP$_{\rm X}$ is the zero-point associated with filter X. We
 have obtained ZP$_{\rm X}$ for ACS/WFC using the ``ACS Zeropoints
 calculator''\footnote{https://acszeropoints.stsci.edu/}; for
 WFC3/UVIS we adopted the zero-points tabulated by
 \citet{2017wfc..rept...14D}.

Table~\ref{tab1} contains the values of $<\delta m> + {\rm ZP}_{\rm X}$ and the
associated errors for each cluster and for each filter.

\begin{figure*}
\includegraphics[width=0.90\textwidth]{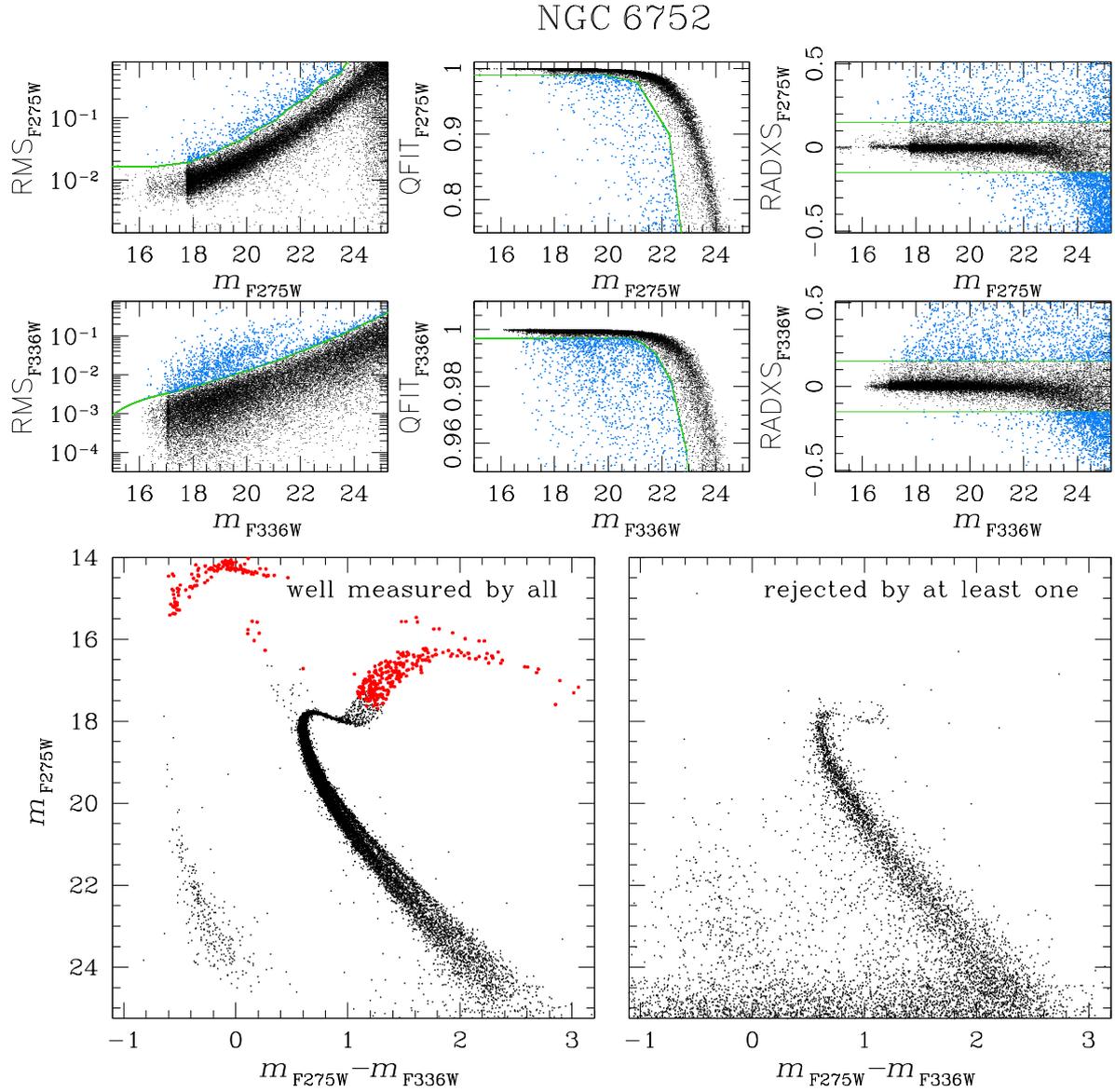}
\caption{Procedure adopted for selecting well-measured stars in the
  cluster NGC\,6752. Top and middle panels show the selection of the
  stars based on \texttt{RMS} (left-hand panels), \texttt{QFIT}
  (central panels), and \texttt{RADXS} (right-hand panels) parameters:
  in azure the stars that are rejected. Bottom panels show the $m_{\rm
    F275W}$ versus $(m_{\rm F275W}-m_{\rm F336W})$ CMD cleaned by
  rejected stars in both filters (left-hand panel) and the CMD of
  the stars rejected in at least one filter. Red dots correspond to 
  the stars that are saturated in at least one filter.  \label{fig:2}}
\end{figure*}

\begin{figure*}
\includegraphics[width=0.95\textwidth]{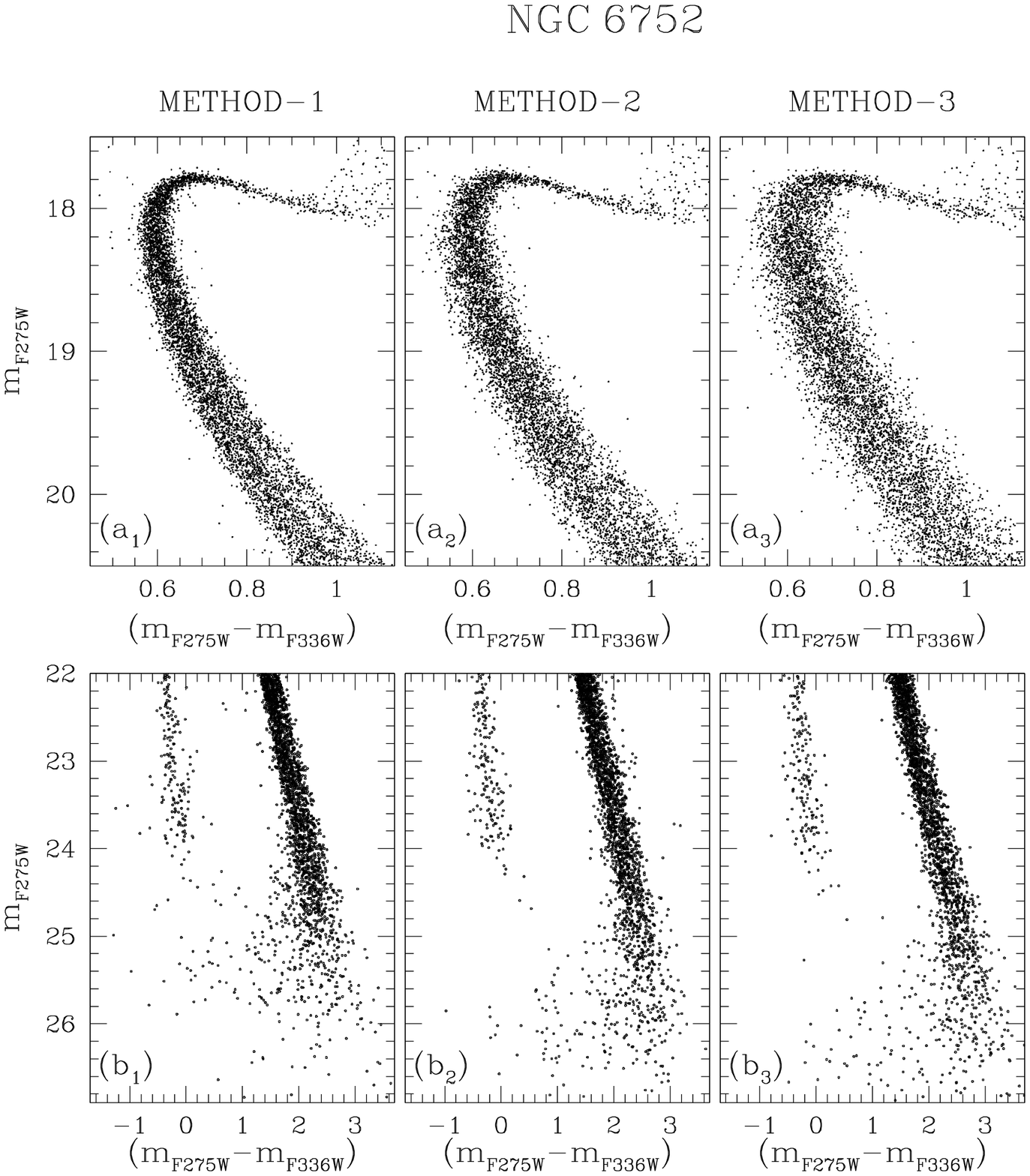}
\includegraphics[bb=17 190 591 373, width=0.95\textwidth]{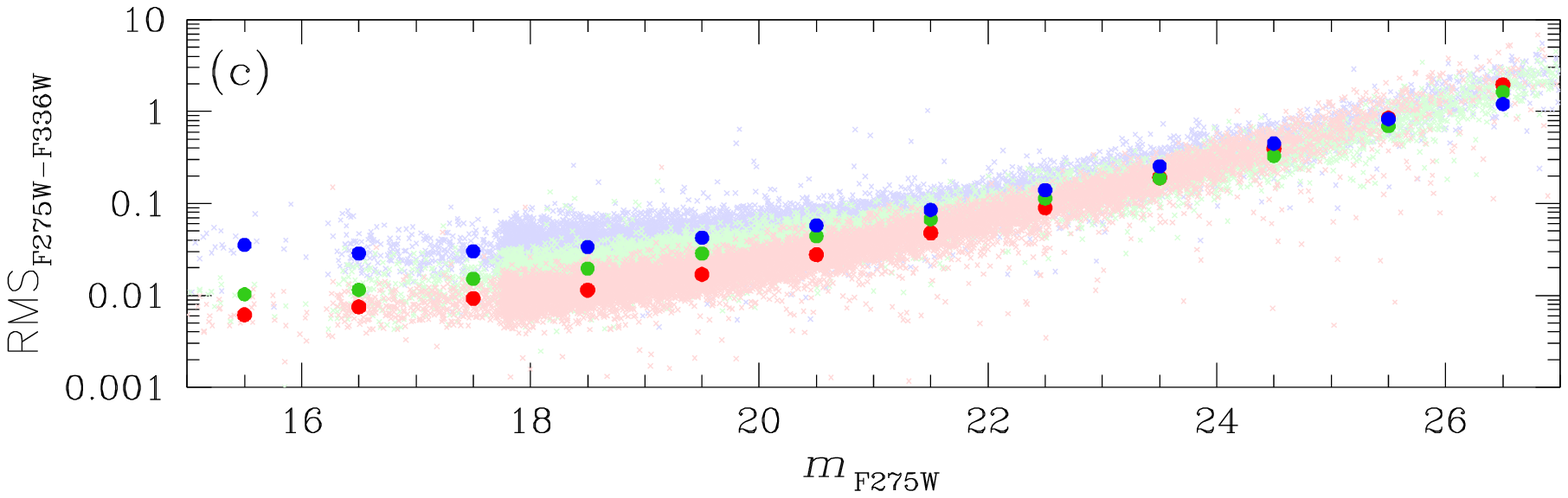}
\caption{The CMDs of NGC\,6752 obtained using the three photometric
  methods described in Sect.~\ref{sec:obs}: top-panels show the
  $m_{\rm F275W}$ versus $(m_{\rm F275W}-m_{\rm F336W})$ CMDs for
  stars with $16.5<m_{\rm F275W}<20.5$ and obtained with method-1
  (panel (a$_1$)), method-2 (panel (a$_2$)), and method-3 (panel
  (a$_3$)).  Middle panels (b$_1$), (b$_2$), and (b$_3$) show the same
  CMDs for stars with $22.0<m_{\rm F275W}<26.9$. Panel (c) shows the
  colour \texttt{RMS} as a function of the F275W magnitude for the
  three photometric methods: in red the Method-1, in green the
  Method-2, and in blue the Method-3.
 \label{fig:3}}
\end{figure*}

\subsection{Astrometric solution}

As described in Section~\ref{ssec:1stp}, our reference frame was based on the {\it Gaia} 
Data Release 1 catalogue (\citealt{2016A&A...595A...2G}), which enables us
to transform the coordinates from (\texttt{X},\texttt{Y}) into 
($\alpha$,$\delta$). As such, the positions are given for Equinox J2000 
and referred to the epoch of {\it Gaia} observations (2015.0).

\subsection{Quality parameters}

In addition to the photometric error (\texttt{RMS}), the KS2 routine
provides as output some quality parameters that are useful for
selecting the best measured stars for a particular science case.

Among them, there is the quality-of-fit (\texttt{QFIT}) parameter that
gives information about the goodness of the PSF-fitting during the
measurement of the position and the flux of a star. It allows us to
distinguish between stars and other sources (of astrophysical nature
or not, e.g., cosmic rays, hot pixels, extended sources, etc). This
parameter is computed using all the pixels of the raster where the
source is measured (pixd), and pixels with neighbour stars are
down-weighted. It is simply the linear-correlation coefficient between
the observed and modelled pixels and is given by:
\begin{equation}
{\tt QFIT} = \frac{\sum\limits_{i,j} {\rm pixd(i,j)}\,{\rm PSF}(i,j)}{\sqrt{\sum\limits_{i,j} {\rm PSF^2}(i,j) \sum\limits_{i,j} {\rm pixd^2}(i,j)}}
\end{equation}
where the sum is performed on a $5\times5$ pixel$^2$ raster pixd
(after neighbour subtraction) centred on the target stars, and
PSF$(i,j)$ is the value of the local PSF-model expected in the pixel
$(i,j)$.

Introduced by \citet[see also
  \citealt{2009ApJ...697..965B,2010ApJ...708L..32B,2018MNRAS.tmp..176S}]{2008ApJ...678.1279B},
the parameter \texttt{RADXS} is a shape parameter that allows us to
distinguish sources that deviate from a PSF shape.  It is a comparison
between the measured flux of the source outside the core (in an
annulus $1.0<r<2.5$\,pixels) and the flux expected from the
PSF-model. For \texttt{RADXS}>0 the source is broader than that
expected from the model (i.e., galaxies), while for negative values of
\texttt{RADXS} the source is sharper than the PSF (i.e., cosmic rays
and artifacts).

Finally, the KS2 routine gives the number of images in which a star
is found (\texttt{N$_{\tt{f}}$}) and the number of consistent measurements 
of the star used to compute its average position and flux 
(\texttt{N$_{\tt{g}}$}).

\section{Colour-magnitude diagrams}
\label{sec:cmds}

In Fig.~\ref{fig:2} we show an example of selection of well-measured
stars for the case of NGC\,6752 and the photometric method-1. Similar
plots can be made for the methods 2 and 3. The top and middle panels
of Fig.~\ref{fig:2} show the selection of the stars based on the
distribution of the photometric errors (\texttt{RMS}, left-hand
panels), the quality of fit (\texttt{QFIT}, center panels), and the
shape of the sources (\texttt{RADXS}, right-hand panels), in the case
of F275W (top panels) and F336W (middle panels) filters. 
The selection based on \texttt{RMS} and \texttt{QFIT} are performed as
done by \cite{2012A&A...540A..16M}: we divided the distributions in 12
magnitude bins and, in each bin, we calculated the 3.5$\sigma$-clipped
average of the magnitude and of the parameter, where $\sigma$ is
  the standard deviation associated to the average value in the given
  bin. We added to the mean parameter of each bin $3.5 \times \sigma$,
  and we linearly interpolated these points (green line).  We excluded all the points above (in
the case of the \texttt{RMS}) or below (in the case of the
\texttt{QFIT})  the green line (azure points). For the sharp parameter, we
selected all the stars that satisfy the condition:
$-0.15<$\texttt{RADXS}$<+0.15$. Bottom panels show the $m_{\rm F275W}$
versus $(m_{\rm F275W}-m_{\rm F336W})$ colour-magnitude diagram (CMD)
for the stars that pass the selection criteria in both filters
(left-hand panel) and for the stars that were rejected in at least one
filter (right-hand panel). In red, the stars that are saturated in at
least one filter and that have been recovered from the ``first-pass''
photometry. From the CMDs, it is clear that many stars with poor
photometric quality are rejected

Figure~\ref{fig:3} shows a comparison of the CMDs of NGC\,6752
obtained using the three photometric methods described in
Sect.~\ref{sec:obs}: top-panels show the $m_{\rm F275W}$ versus
$(m_{\rm F275W}-m_{\rm F336W})$ CMDs of NGC\,6752 in the bright regime
of magnitudes ($16.5<m_{\rm F275W}<20.5$) and obtained with method-1
(panel (a$_1$)), method-2 (panel (a$_2$)), and method-3 (panel
(a$_3$)).  The middle panels (b$_1$), (b$_2$), and (b$_3$) of
Fig.~\ref{fig:3} show the same CMDs for stars with $22.0<m_{\rm
  F275W}<26.9$. The stars plotted in the CMDs have passed the
selection criteria above described applied to each photometric
method. Top panels show that for bright stars method-1 gives better
results than method-2 and method-3. Panel (b$_1$) shows that method-1
gives a good measurement of the stars with $m_{\rm F275W}\lesssim
23.0$; stars having magnitude $23.0\lesssim m_{\rm F275W}\lesssim
25.5$ are well measured with method-2, while method-3 is an optimal
choice for stars having $25.5\lesssim m_{\rm F275W}\lesssim27.0$.
Panel (c) of Fig.~\ref{fig:3} shows the $m_{\rm F275W}-m_{\rm F336W}$
colour \texttt{RMS} as a function of the F275W magnitude: light red,
light green, and light blue points are the \texttt{RMS} of the stars
measured with method-1, method-2, and method-3, respectively. We
divided the \texttt{RMS} distribution in bins of width 1 F275W
magnitude, and we computed in each bin the median \texttt{RMS}. The
binned \texttt{RMS} distributions of the three methods (in red, green,
and blue for methods 1, 2, and 3, respectively), confirm that method-1
gives the best results in the bright magnitude regime, while stars
measured with methods-2 and 3 have lower \texttt{RMS} at fainter
magnitudes.

\begin{figure}
\includegraphics[width=0.5\textwidth, bb=20 147 305 701]{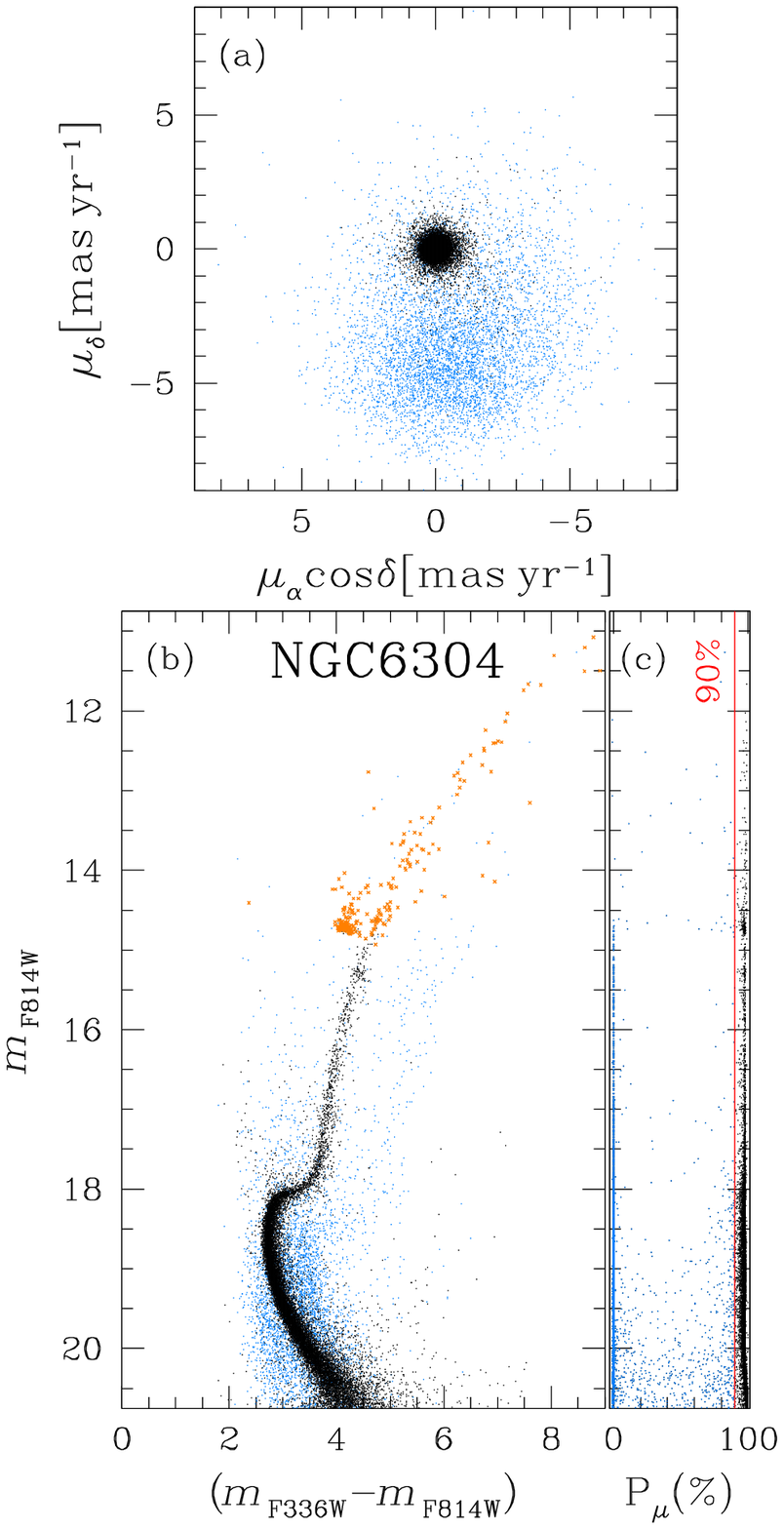}
\caption{Likely NGC\,6304 members selection. Panel (a) shows the VPD
  for the well-measured stars; panel (b) illustrates the $m_{\rm
    F814W}$ vs. $m_{\rm F336W}-m_{\rm F814W}$ CMD: in orange are the
  saturated stars in at least one filter; panel (c) is the membership
  probability as a function of the F814W magnitude. In all the plots
  we highlighted in azure the stars having \texttt{P}$_\mu<90$\,\% and in black
  the likely cluster members. \label{fig:4}}
\end{figure}

\begin{table*}
  \caption{Epochs of observations}
    \label{tab2}
    \begin{tabular}{ l c c c l c c c}
\hline
\multicolumn{1}{c}{Cluster} &
\multicolumn{1}{c}{1st epoch} &
\multicolumn{1}{c}{2nd epoch} &
\multicolumn{1}{c}{$\Delta$t (yrs)} &
\multicolumn{1}{c}{Cluster} &
\multicolumn{1}{c}{1st epoch} &
\multicolumn{1}{c}{2nd epoch} &
\multicolumn{1}{c}{$\Delta$t (yrs)} \\
\hline
   NGC0104  &   2006.20    &   2013.14     &     6.94  &    NGC6352  &   2006.27    &   2014.01     &     7.74  \\
   NGC0288  &   2006.56    &   2012.83     &     6.27  &    NGC6362  &   2006.41    &   2014.37     &     7.96  \\
   NGC0362  &   2006.42    &   2012.70     &     6.28  &    NGC6366  &   2006.25    &   2014.50     &     8.25  \\
   NGC1261  &   2006.19    &   2013.93     &     7.74  &    NGC6388  &   2006.27    &   2014.44     &     8.18  \\
   NGC1851  &   2006.33    &   2014.51     &     8.17  &    NGC6397  &   2006.41    &   2014.34     &     7.93  \\
   NGC2298  &   2006.45    &   2014.27     &     7.82  &    NGC6441  &   2006.41    &   2014.36     &     7.95  \\
   NGC2808  &   2006.17    &   2013.69     &     7.52  &    NGC6496  &   2006.31    &   2014.01     &     7.71  \\
   NGC3201  &   2006.20    &   2013.85     &     7.65  &    NGC6535  &   2006.25    &   2014.48     &     8.23  \\
   NGC4590  &   2006.18    &   2014.07     &     7.88  &    NGC6541  &   2006.25    &   2014.29     &     8.04  \\
   NGC4833  &   2006.57    &   2014.16     &     7.59  &    NGC6584  &   2006.40    &   2014.02     &     7.62  \\
   NGC5024  &   2006.17    &   2014.04     &     7.87  &    NGC6624  &   2006.29    &   2014.08     &     7.79  \\
   NGC5053  &   2006.18    &   2014.15     &     7.97  &    NGC6637  &   2006.39    &   2014.32     &     7.93  \\
   NGC5272  &   2006.14    &   2012.37     &     6.23  &    NGC6652  &   2006.40    &   2013.93     &     7.52  \\
   NGC5286  &   2006.17    &   2013.95     &     7.78  &    NGC6656  &   2006.25    &   2014.54     &     8.29  \\
   NGC5466  &   2006.28    &   2014.13     &     7.85  &    NGC6681  &   2006.39    &   2014.09     &     7.70  \\
   NGC5897  &   2006.27    &   2014.25     &     7.97  &    NGC6715  &   2006.40    &   2014.09     &     7.69  \\
   NGC5904  &   2006.20    &   2014.31     &     8.11  &    NGC6717  &   2006.24    &   2014.45     &     8.21  \\
   NGC5927  &   2006.28    &   2014.63     &     8.34  &    NGC6723  &   2006.42    &   2014.41     &     7.99  \\
   NGC5986  &   2006.29    &   2014.60     &     8.31  &    NGC6752  &   2006.40    &   2010.34     &     3.95  \\
   NGC6093  &   2006.27    &   2012.44     &     6.17  &    NGC6779  &   2006.36    &   2014.17     &     7.81  \\
   NGC6101  &   2006.42    &   2014.19     &     7.77  &    NGC6791  &   2004.74    &   2013.97     &     9.23  \\
   NGC6121  &   2006.18    &   2014.82     &     8.65  &    NGC6809  &   2006.30    &   2014.44     &     8.14  \\
   NGC6144  &   2006.29    &   2014.28     &     7.99  &    NGC6838  &   2006.36    &   2014.08     &     7.71  \\
   NGC6171  &   2006.25    &   2014.32     &     8.08  &    NGC6934  &   2006.25    &   2014.20     &     7.95  \\
   NGC6205  &   2006.25    &   2012.37     &     6.12  &    NGC6981  &   2006.38    &   2014.11     &     7.73  \\
   NGC6218  &   2006.17    &   2014.01     &     7.85  &    NGC7078  &   2006.33    &   2011.79     &     5.46  \\
   NGC6254  &   2006.18    &   2014.02     &     7.84  &    NGC7089  &   2006.29    &   2013.69     &     7.40  \\
   NGC6304  &   2006.29    &   2014.04     &     7.76  &    NGC7099  &   2006.34    &   2014.54     &     8.20  \\
   NGC6341  &   2006.28    &   2014.20     &     7.92  &             &              &               &           \\
\hline
\end{tabular}

\end{table*}

\section{Relative Proper Motions and cluster Membership probabilities}
\label{sec:rpm}

The photometric catalogs published with this manuscript are obtained
with reduction pipelines fine tuned to achieve high-precision
photometric measurements. High-precision proper motions require a
completely different, ad-hoc reduction of the images (see, e.g., \citealt{2014ApJ...797..115B,2018ApJ...853...86B,2018arXiv180505332L}),
which is beyond the scope of the present manuscript.
High-precision astrometry have different demands with respect to
high-precision photometry since, to the first order, photometry cares
of sum of pixels while astrometry is focused on differences between
pixels.
There are several systematic effects (e.g., CTE correction residuals,
geometric-distortion correction residuals, color terms in the blue
filters of the WFC3/UVIS, etc.) that cannot be properly accounted for
with a reduction fine-tuned for photometry. These systematic effects
can be as large as 0.2 ACS/WFC pixels in a given data set.  The proper
motions we computed for this work are based on image pairs, are
insensitive to systematic errors, and are highly degenerate in terms of
proper motion errors.
 
In summary, the present proper motions can be only used to calculate
cluster membership in order to separate cluster members and field
stars. We are presently working on the much more precise astrometry
needed for internal kinematics ($\ll$0.1\,mas/yr, see,
e.g. \citealt{2018arXiv180505332L}), and we  defer the publication
of proper motions catalogues to future papers.

Selecting bona-fide cluster members by relying solely on the stellar
positions on a CMD is not an easy task, in particular for those GCs
near the Galactic plane or bulge. In principle, the user can combine
our photometry with the proper motions in the \textit{Gaia} Data Release
2 \citep{2016A&A...595A...1G,2018arXiv180409365G}.  However, the
\textit{Gaia} catalogue is severely incomplete near the core of GCs
(see, e.g., \citealt{2018arXiv180505332L}), and furthermore most 
cluster stars are well below Gaia's faint limit.  Therefore, in order to
help interested users to select cluster members, we include in our
photometric catalogues an estimate of the membership probability
\texttt{P}$_\mu$. In this section we will describe how we measured relative
motions and estimated membership probabilities.

To compute relative PMs, we adopted the approach described in many
previous publications by our group (see, e.g., \citealt{2003AJ....126..247B,
2006A&A...454.1029A,2008A&A...484..609Y,2010A&A...517A..34B,
2014A&A...563A..80L,2015A&A...573A..70N,2015MNRAS.450.1664L,
2016MNRAS.455.2337N,2018ApJ...853...15K}).
The routine KS2 provides raw catalogues, one for each exposure,
containing positions and magnitudes of the stars listed in the final
catalogue as measured on the single images. We used these raw
catalogues to compute the relative PMs; for this computation we
excluded F275W raw catalogues because of colour-dependent systematic
effects in the geometric-distortion correction of this filter (\citealt{2011PASP..123..622B})

We used six-parameter local transformations and a sample of likely
cluster members (red-giant branch, RGB, sub-giant branch, SGB, and main sequence, MS, stars) to compute the displacement
between the stellar positions in two different epochs.  We started
with a first, preliminary, sample of likely cluster members, selected
on the $m_{\rm F814W}$ vs. $(m_{\rm F606W}-m_{F814W})$ CMD, to compute
the coefficients of the six-parameter linear transformations between
the positions of the raw catalogues and the final catalogue. In order to
minimise the effects of residual uncorrected geometric distortion, we
computed the transformations using local samples (50 stars) of likely
cluster members. Stars in each single-exposure catalogue of
the first-epoch data set were compared to stars in each
single-exposure catalogue of the second-epoch data set. Suppose we have
$N$ exposures for the first epoch and $M$ exposures for the second
epoch, then we end up with $N\times M$ displacements for each
star. The computed relative proper motion of a star is the average of
all these displacements along the X and the Y axes. The assigned error
is simply the RMS of the displacement residuals around the
average. Because the displacements are not statistically independent,
the assigned errors are not a reliable estimate of the proper-motion
errors, but can still be used to estimate membership probabilities
(see \citealt{2006A&A...454.1029A} for an in-depth description of the
method). We used these displacements to remove from
the list of likely cluster members objects that had colours placing them 
close to the cluster sequences but had a field-star-like motion (i.e. those
stars with proper motions relative to the cluster mean motion
$>6$\,mas\,yr$^{-1}$). We iterated the procedure three times using
the new member list to compute the improved linear transformations with
each iteration.

KS2 does not measure the positions and fluxes of saturated stars.
Therefore, we used the outputs of first pass photometry to obtain the
relative proper motions of these stars.

Since the coefficients of the six-parameter linear transformations
are computed using likely cluster members, the stellar displacements
are computed relative to the cluster mean motion, and therefore, in
the vector-point diagram (VPD), cluster stars will be centred around
(0,0), while field stars will lie in different regions of the VPD.
The mean date of the adopted observations for the first and second
epoch and the time baseline are listed in
Table~\ref{tab2}.

Membership probabilities were then computed using the local-sample
method, similarly to what was done in \citet{2009A&A...493..959B} and
\citet{2014A&A...563A..80L}. For each target star, the membership
probability is estimated using a sub-sample of reference 500 stars in
the catalogue. These reference stars were initially chosen on the basis
of PM error (typically $\pm 0.25$ mas yr$^{-1}$) and a magnitude
similar to those of the target. The only exceptions are for target
stars along the SGB and RGB, for which---due to small-number
statistics---we considered as reference stars sources over the
entire SGB-RGB sequence.

The cluster density function is modelled with an axisymmetric 2D
Gaussian distribution centred on the origin of the VPD (since PMs are
computed relative to the cluster's bulk motion). The sigma of the 2D
Gaussian is magnitude dependent, and is defined as the 68.27$^{\rm
  th}$ percentile of the $\sqrt{\mu_x^2+\mu_y^2}$ distribution at any
given magnitude.
Field stars are assumed to have a flat distribution in the VPD, which
is a fair assumption for the vast majority or our clusters. The
remaining parameters of the local-sample method \citep[see Eq.~10
  of][]{1995AJ....109..672K} are solved-for using least-squares
techniques.

Figure~\ref{fig:4} shows an example of field-star decontamination
based on membership probabilities. Panels (a), (b), and (c) show the
VPD, the $m_{\rm F814W}$ vs. $m_{\rm F336W}-m_{\rm F814W}$ CMD, and
the membership distribution \texttt{P}$_\mu$, respectively, for all the
well-measured stars of NGC\,6304: in black are the stars having a
membership probability \texttt{P}$_\mu>90$\,\%, in azure the other stars. In
panel (b) we highlight in orange the stars that are saturated in at
least one of the two filters.

Stars with unrealistic PM errors\footnote{Stars with extremely
  (several sigmas) underestimated or overestimated PM errors with
  respect to those of stars with similar magnitude.} are not
considered in our membership-probability determination. This limits
our ability to estimate membership-probabilities to stars brighter
than a certain magnitude threshold that varies from cluster to
cluster.

\begin{figure*}
\includegraphics[width=0.95\textwidth]{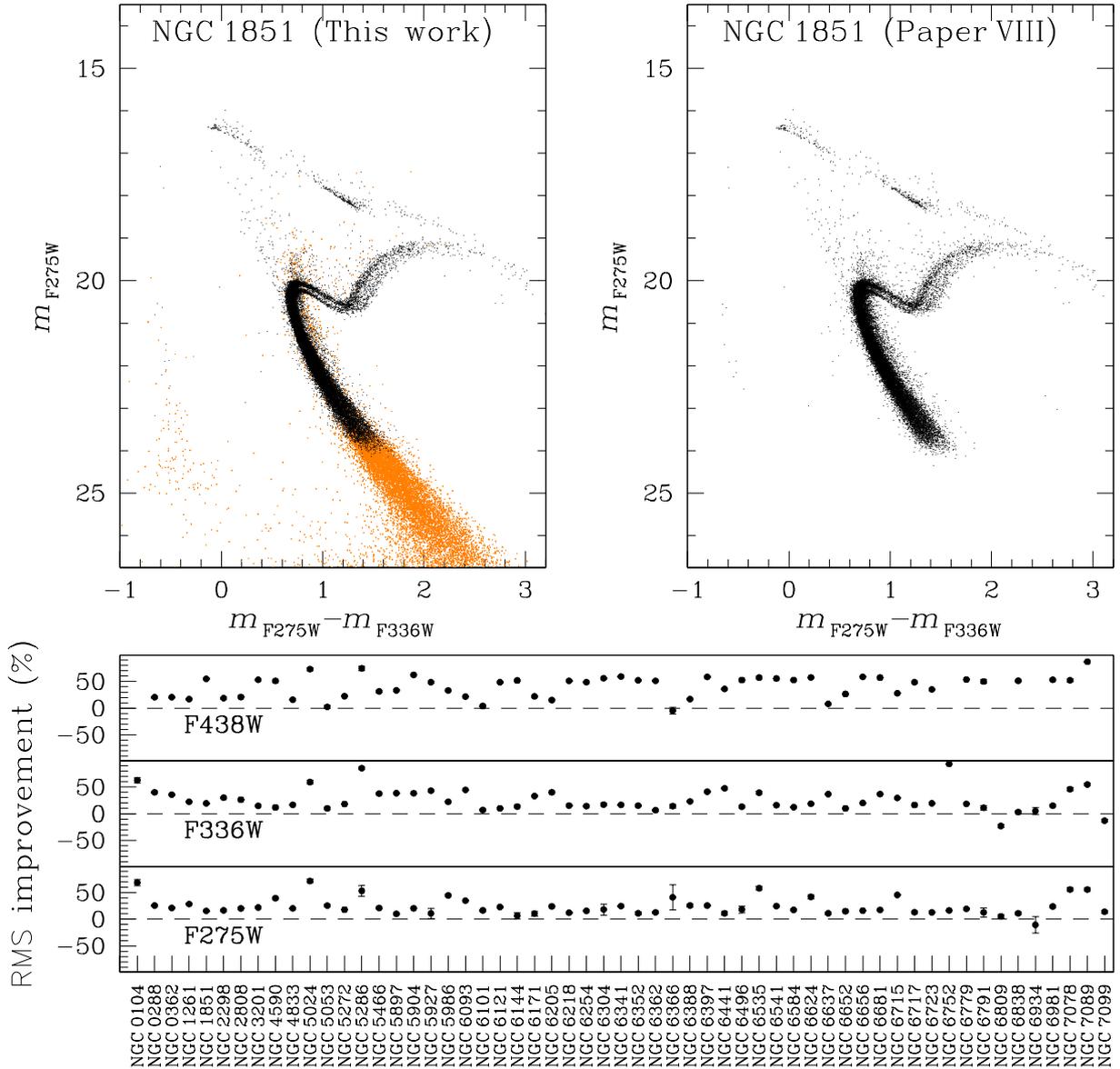}
\caption{
Comparison between catalogues of present work and
\citetalias{2017AJ....153...19S}. Top-panels show the $m_{\rm F275W}$
vs. $m_{\rm F275W}-m_{\rm F336W}$ CMDs of NGC\,1851 obtained with the
catalogue obtained in the present work (left-hand panel) and the
\citetalias{2017AJ....153...19S} catalogue (right-hand panel). Black points are the stars
in common to the two data-release, orange points the stars measured in
this work. Bottom panels show the photometric \texttt{RMS}
improvements of our catalogues with respect to
\citetalias{2017AJ....153...19S} ones for F275W, F336W, and F438W
bands.
 \label{fig:5}}
\end{figure*}

\section{The need for a new data release}
\label{sec:newdr}
In \citetalias{2017AJ....153...19S} preliminary catalogues of the
clusters in our project were released, in order to provide to
the astronomical community an initial estimate of positions, 
luminosities, and colours of bright stars belonging to different 
populations in order to enable target selection for spectroscopic 
observations. As clearly stated in \citetalias{2017AJ....153...19S}, 
this was the main purpose of the preliminary published catalogues.

The data-reduction pipelines used in this work and in
\citetalias{2017AJ....153...19S} are different.
In the following, we list the most important improvements.
\begin{description}
\item [{\bf 1) Perturbed PSFs:}] In \citetalias{2017AJ....153...19S}
  static library PSFs were used. As explained in Sect.~\ref{sec:obs},
  in this work we perturbed library PSFs to take into account of
  spatial and temporal variations of the PSFs and to empirically
  reproduce the shape of the stars in each single image.  This
  procedure was not adopted in \citetalias{2017AJ....153...19S};
\item [{\bf 2) Neighbour subtraction:}] For the present catalogue,
  when we measured the position and the flux of each source, we
  subtracted the neighbours to avoid the contamination by other close
  stars. This allowed us to better estimate the real flux of each
  star (as well as the measurement error), even in very crowded 
  environments.  In \citetalias{2017AJ....153...19S} neighbour 
  stars were not subtracted;
\item [{\bf 3) Faint stars:}] Because in
  \citetalias{2017AJ....153...19S} we were interested only in measuring
  bright stars, only stars with S/N$\gtrsim 10$ were searched in each
  image. The main consequence is that the faint part of the CMD was 
  lacking in \citetalias{2017AJ....153...19S}.  In the present work we
  searched for each significant peak ($\geq 1\sigma$ above the sky)
  combining all the images, and measured the associated source using
  three different photometric methods.

\item [{\bf 4) Optical filters and UV completeness:}] In
  \citetalias{2017AJ....153...19S} UV starlists were cross-identified
  with former ACS\,GCS catalogues (\citealt{2007AJ....133.1658S}), and
  many bright stars in UV bands were lost. In the present work we also
  re-reduced data from the GO-10775 in an effort to improve the photometry 
  in F606W and F814W bands using the new pipeline. Moreover, because we
  searched for stars using all filters, the new catalogues include
  stars bright in UV, even if they are too faint to be detected in
  optical bands (e.g., white dwarfs).
\end{description}

\begin{table*}
  \caption{Description of the column content of the astro-photometric catalogues.}
    \label{tab3}
    \begin{tabular}{ l c c l}
\hline
\multicolumn{1}{c}{Column} &
\multicolumn{1}{c}{Name} &
\multicolumn{1}{c}{Unit} &
\multicolumn{1}{c}{Explanation} \\
\hline
01,02  & \texttt{X}, \texttt{Y}   & [pix]  &  (x,y) stellar position in a reference system where the cluster center is in (5000,5000)   \\
03  & $m_{\rm F275W}$               & [mag]  &  F275W calibrated magnitude  \\
04  & \texttt{RMS}$_{\rm F275W}$    & [mag]  &  F275W photometric RMS  \\
05  & \texttt{QFIT}$_{\rm F275W}$   &   &  F275W quality-fit parameter  \\
06  & \texttt{RADXS}$_{\rm F275W}$  &   &  F275W sharp parameter  \\
07  & \texttt{N}$_{{\tt f},{\rm F275W}}$  &    &  Number of F275W exposures the source is found [99: saturated star]  \\
08  & \texttt{N}$_{{\tt g},{\rm F275W}}$  &    &  Number of F275W exposures the source is well measured [99: saturated star]  \\
09  & $m_{\rm F336W}$                  & [mag] &     F336W calibrated magnitude  \\
10  & \texttt{RMS}$_{\rm F336W}$       & [mag] &     F336W photometric RMS  \\
11  & \texttt{QFIT}$_{\rm F336W}$      &  &     F336W quality-fit parameter  \\
12  & \texttt{RADXS}$_{\rm F336W}$     &  &     F336W sharp parameter  \\
13  & \texttt{N}$_{{\tt f},{\rm F336W}}$  &    &  Number of F336W exposures the source is found [99: saturated star]  \\
14  & \texttt{N}$_{{\tt g},{\rm F336W}}$  &     &  Number of F336W exposures the source is well measured [99: saturated star]  \\
15  & $m_{\rm F438W}$                  & [mag] &     F438W calibrated magnitude  \\
16  & \texttt{RMS}$_{\rm F438W}$       & [mag] &     F438W photometric RMS  \\
17  & \texttt{QFIT}$_{\rm F438W}$      &  &     F438W quality-fit parameter  \\
18  & \texttt{RADXS}$_{\rm F438W}$     &  &     F438W sharp parameter  \\
19  & \texttt{N}$_{{\tt f},{\rm F438W}}$  &     &  Number of F438W exposures the source is found [99: saturated star]  \\
20  & \texttt{N}$_{{\tt g},{\rm F438W}}$  &     &  Number of F438W exposures the source is well measured [99: saturated star]  \\
21  & $m_{\rm F606W}$                  & [mag] &     F606W calibrated magnitude  \\
22  & \texttt{RMS}$_{\rm F606W}$       & [mag] &     F606W photometric RMS  \\
23  & \texttt{QFIT}$_{\rm F606W}$      &  &     F606W quality-fit parameter  \\
24  & \texttt{RADXS}$_{\rm F606W}$     &  &     F606W sharp parameter  \\
25  & \texttt{N}$_{{\tt f},{\rm F606W}}$  &     &  Number of F606W exposures the source is found [99: saturated star]  \\
26  & \texttt{N}$_{{\tt g},{\rm F606W}}$  &     &  Number of F606W exposures the source is well measured [99: saturated star]  \\
27  & $m_{\rm F814W}$                  & [mag] &     F814W calibrated magnitude  \\
28  & \texttt{RMS}$_{\rm F814W}$       & [mag] &     F814W photometric RMS  \\
29  & \texttt{QFIT}$_{\rm F814W}$      &  &     F814W quality-fit parameter  \\
30  & \texttt{RADXS}$_{\rm F814W}$     &  &     F814W sharp parameter  \\
31  & \texttt{N}$_{{\tt f},{\rm F814W}}$  &     &  Number of F814W exposures the source is found [99: saturated star]  \\
32  & \texttt{N}$_{{\tt g},{\rm F814W}}$  &     &  Number of F814W exposures the source is well measured [99: saturated star]  \\
33  & \texttt{P$_{\mu}$}                    &  [\%] & Membership probability [-1.0: not available] \\
34  & $\alpha$                       &  [deg.] & Right ascension (J2000, epoch 2015) of the star \\
35  & $\delta$                       &  [deg.] & Declination (J2000, epoch 2015) of the star \\
36  & \texttt{ID}                    &         &  Identification number of the star \\
37  & \texttt{ITER}                  &         & Iteration the star was found \\
    &     & &1-5: found in F814W and F606W images  \\
    &     & &  6: found in F438W  images  \\
    &     & &  7: found in F336W  images  \\
    &     & &  8: found in F275W  images  \\
\hline
\end{tabular}

{\bf Note}: For NGC\,0104 and NGC\,6752, the F438W quantities are referred to the ACS/WFC F435W filter.

\end{table*}

Figure~\ref{fig:5} gives an example of the photometric improvements of
the catalogues released by this work with respect to the preliminary
catalogue in \citetalias{2017AJ....153...19S}. The bottom panels show the
\texttt{RMS} improvements (in percentage) of our photometry compared
to that published in \citetalias{2017AJ....153...19S} for the filters
F275W, F336W, and F438W.  The \texttt{RMS} was calculated for the
stars in common between the two catalogues in the magnitude range
$15\leq m_{\rm X} \leq 20$, with X=F275W, F336W, F438W. In this
interval we computed the 3.5$\sigma$-clipped median and dispersion of
\texttt{RMS}$_{\rm X}$ for both catalogues and calculated the value
100$\times$[$1-$\texttt{RMS$_{\rm X}$}(this work)/\texttt{RMS$_{\rm X}$}(\citetalias{2017AJ....153...19S})]
that we used as indicator of photometric
improvement.  On average, our photometry has a $\sim$20-30\% lower
\texttt{RMS} than that published in \citetalias{2017AJ....153...19S}.
Top panels of Fig.~\ref{fig:5} illustrate a comparison between the
$m_{\rm F275W}$ vs. $m_{\rm F275W}-m_{\rm F336W}$ CMDs of NGC\,1851
from method-1 photometry (left panel) and the catalogue published in
\citetalias{2017AJ....153...19S}. In black we show the stars in common
between these catalogues, in orange the stars measured in the present
work, but missing in \citetalias{2017AJ....153...19S} catalogue.  The
photometric improvement is evident, especially at the SGB and MS level.

Previous papers of the series are based on the catalogues described in
\citetalias{2015AJ....149...91P}, which were generated for internal 
use, and have not been published. Even though the routines used to obtain 
may be slightly different from the ones we adopted for the present 
paper, these catalogues were extracted using perturbed PSFs and 
neighbour subtraction.  The F275W, F336W, and F438W photometric 
precision of this dataset and the internal-use set are 
comparable. The main difference regards the optical filters: as with
the preliminary catalogue, the UV catalogues extracted in 
\citetalias{2015AJ....149...91P} were cross-identified with pre-existing
ACS\,GCS catalogues (\citealt{2007AJ....133.1658S}) with all the
limitations we discussed above for UV-bright sources.  The catalogues 
we publish in this paper includes a new reduction of ACS GO-10775 F606W 
and F814W data, and includes UV and optical magnitudes of sources 
detected significantly in at least one of the F275W, F336W, F438W, F606W, 
and F814W bands.

\section{The Data release}
\label{sec:dr}
This new data release replaces the preliminary public available data
release of \citetalias{2017AJ....153...19S} (see
Section~\ref{sec:newdr}). The new released material
  is part of the project ``HST UV Globular cluster Survey''
  (HUGS). All of the data products from HUGS are available at Mikulski
  Archive for Space Telescopes (MAST, \url{http://dx.doi.org/10.17909/T9810F}) as a High Level Science
  Product.

We release the astro-photometric catalogues for all 57 clusters and,
for each of them, we also release all the astrometrised stacked images
(see Sect.~\ref{sec:obs} for details).  The released material will be
available at the ``Exoplanets and Stellar Populations Group'' (ESPG)
website
of the Universit\`a degli Studi di Padova, and on the MAST under the project
  HUGS\footnote{\url{https://archive.stsci.edu/prepds/hugs/}}.

For each cluster we release three catalogues, one for each photometric
method. The catalogues contain information on the positions and on the
photometry of each star found in the field. 
The catalogues also include membership probability.
In Table~\ref{tab3}
we describe the content of each
column. The same description is also included in the header of each catalogue.
For exemplification purpose,  Table~\ref{tab4} show three rows of one of
the released tables.

The catalogues that we make public here are complemented by the
astrometric and photometric catalogues of the external ACS/WFC fields
for 48 GCs plus NGC\,6791 observed in parallel to the GO-13297
WFC3/UVIS central fields and published in
\citetalias{2018MNRAS.tmp..176S}.  All catalogues are available at ESPG webpage

\section*{Acknowledgements}

This work has made use of data from the European Space Agency (ESA)
mission Gaia (http://www.cosmos.esa.int/gaia), processed by the Gaia
Data Processing and Analysis Consortium (DPAC,
http://www.cosmos.esa.int/web/gaia/dpac/consortium).  Funding for the
DPAC has been provided by national institutions, in particular the
institutions participating in the Gaia Multilateral Agreement.  DN and
GP acknowledge partial support by the Universit\`a degli Studi di
Padova Progetto di Ateneo CPDA141214 and BIRD178590 and by INAF under
the program PRIN-INAF2014.  ML and AB acknowledge support from STScI
grant GO~13297.  AA, SC, and GP acknowledge partial support by the
Spanish Ministry of Economy and Competitiveness and the Spanish
Ministry of Science, Innovation and Universities (grants
AYA2013-42781-P and AYA2017-89841-P). AA acknowledges partial support
by the Instituto de Astrof\`isica de Canarias (grant 310394). APM
acknowledges support by the European Research Council through the
ERC-StG 2016 project 716082 'GALFOR'. AFM has been supported by the
Australian Research Council through Discovery Early Career Researcher
Award DE160100851.

\begin{landscape}
\begin{table}
  \caption{Three lines from the catalogue of NGC\,6304.}
    \label{tab4}
    \begin{tabular}{ c c c c c c c c c c c c c c}
\hline
\multicolumn{1}{c}{\bf \texttt{X}                            }  &
\multicolumn{1}{c}{\bf \texttt{Y}                            }  &
\multicolumn{1}{c}{\bf $m_{\rm F275W}$                        }  &
\multicolumn{1}{c}{\bf \texttt{RMS}$_{\rm F275W}$            }  &
\multicolumn{1}{c}{\bf \texttt{QFIT}$_{\rm F275W}$           }  &
\multicolumn{1}{c}{\bf \texttt{RADXS}$_{\rm F275W}$          }  &
\multicolumn{1}{c}{\bf \texttt{N}$_{{\tt f},{\rm F275W}}$    }  &
\multicolumn{1}{c}{\bf \texttt{N}$_{{\tt g},{\rm F275W}}$    }  &
\multicolumn{1}{c}{\bf $m_{\rm F336W}$                       }  &
\multicolumn{1}{c}{\bf \texttt{RMS}$_{\rm F336W}$            }  &
\multicolumn{1}{c}{\bf \texttt{QFIT}$_{\rm F336W}$           }  &
\multicolumn{1}{c}{\bf \texttt{RADXS}$_{\rm F336W}$          }  &
\multicolumn{1}{c}{\bf \texttt{N}$_{{\tt f},{\rm F336W}}$    }  &
\multicolumn{1}{c}{\bf \texttt{N}$_{{\tt g},{\rm F336W}}$    }  \\
\hline
6993.4438 &  2488.7637  &    24.0801  &    0.0000  &    0.7845  &   -0.4151 & 1 & 1  &    22.3862  &    0.0137   &   0.9844  &    0.0308 & 2 & 1\\
4117.0439 &  2489.3533  &    23.1821  &    0.0000  &    0.9453  &   -0.0596 & 1 & 1  &    21.4650  &    0.0103   &   0.9955  &    0.1101 & 2 & 1\\
4674.6870 &  2490.8945  &   -99.9999  &   99.9999  &    0.0000  &    9.9999 & 0 & 0  &   -99.9999  &   99.9999   &   0.0000  &    9.9999 & 0 & 0\\
\hline
\end{tabular}

\vspace{1cm}

\begin{tabular}{ c c c c c c c c c c c c c c }
\hline
\multicolumn{1}{c}{\bf $m_{\rm F438W}$                        }  &
\multicolumn{1}{c}{\bf \texttt{RMS}$_{\rm F438W}$            }  &
\multicolumn{1}{c}{\bf \texttt{QFIT}$_{\rm F438W}$           }  &
\multicolumn{1}{c}{\bf \texttt{RADXS}$_{\rm F438W}$          }  &
\multicolumn{1}{c}{\bf \texttt{N}$_{{\tt f},{\rm F438W}}$    }  &
\multicolumn{1}{c}{\bf \texttt{N}$_{{\tt g},{\rm F438W}}$    }  &
\multicolumn{1}{c}{\bf $m_{\rm F606W}$                       }  &
\multicolumn{1}{c}{\bf \texttt{RMS}$_{\rm F606W}$            }  &
\multicolumn{1}{c}{\bf \texttt{QFIT}$_{\rm F606W}$           }  &
\multicolumn{1}{c}{\bf \texttt{RADXS}$_{\rm F606W}$          }  &
\multicolumn{1}{c}{\bf \texttt{N}$_{{\tt f},{\rm F606W}}$    }  &
\multicolumn{1}{c}{\bf \texttt{N}$_{{\tt g},{\rm F606W}}$    }  &
\multicolumn{1}{c}{\bf  $m_{\rm F814W}$                       }  &
\multicolumn{1}{c}{\bf  \texttt{RMS}$_{\rm F814W}$            }  \\
\hline
21.3648     &   0.0000     &   0.9933     &   0.1235  & 1   & 1     &   19.5705    &    0.0025      &  1.0000    &    0.0013   & 4   & 1  &  18.4205     &   0.0027   \\
21.1103     &   0.0000     &   0.9960     &  -0.0177  & 1   & 1     &   19.7744    &    0.0035      &  0.9999    &    0.0324   & 4   & 1  &  18.7334     &   0.0032   \\
-99.9999    &   99.9999    &    0.0000    &    9.9999 & 0   & 0     &   19.9864    &    0.0000      &  0.9999    &    0.0615   & 1   & 1  &  18.9580     &   0.0000   \\
\hline
\end{tabular}

\vspace{1cm}

\begin{tabular}{ c c c c c c c c c c     }
\hline
\multicolumn{1}{c}{\bf  \texttt{QFIT}$_{\rm F814W}$           }  &
\multicolumn{1}{c}{\bf  \texttt{RADXS}$_{\rm F814W}$          }  &
\multicolumn{1}{c}{\bf  \texttt{N}$_{{\tt f},{\rm F814W}}$    }  &
\multicolumn{1}{c}{\bf  \texttt{N}$_{{\tt g},{\rm F814W}}$    }  &
\multicolumn{1}{c}{\bf \texttt{P$_\mu$}} &
\multicolumn{1}{c}{\bf $\alpha$} &
\multicolumn{1}{c}{\bf $\delta$} &
\multicolumn{1}{c}{\bf \texttt{ID}} &
\multicolumn{1}{c}{\bf \texttt{ITER}} \\ 
\hline
        1.0000     &   0.0049   & 4   & 1 &  00.0 &           258.638467 &          -29.489565 &   R0000287 &   1\\ 
        1.0000     &   0.0595   & 4   & 1 &  98.1 &           258.638913 &          -29.489539 &   R0000288 &   1\\ 
        0.9999     &   0.0999   & 1   & 1 & -01.0 &           258.614414 &          -29.489541 &   R0000289 &   1\\ 
\hline
\end{tabular}

\end{table}
\end{landscape}

\bibliographystyle{mnras}
\bibliography{biblio} 

\bsp	
\label{lastpage}
\end{document}